\documentclass[aps,twocolumn]{revtex4}
\usepackage{epsfig}
\usepackage{amsmath}
\usepackage{epsfig}

\begin{document}

\title
{\bf Bound state eigenvalues from transmission coefficients}    
\author{Zafar Ahmed$^{1,2}$ and Koushik Bhattacharya$^3$}
\email{zai-alpha@hotmail.com, itskb30121@gmail.com}
\affiliation{$~^{1}$Nuclear Physics Division, Bhabha Atomic Research Cntre,Mumbai 400 085, India;\\  $~^2$Homi Bhabha National Institute, Anushaktinagar, Mumbai, 400094, India \\$~^3$Human Resource Development Division, Bhabha Atomic Research Centre,Mumbai 400 094, India}
\date{\today}
\begin{abstract}
Experts know that bound state energy eigenvalues of a potential well are  poles of its transmission amplitude, $t(E)$. The textbooks do well by solving the bound states and scattering states separately, but the connection goes unheeded.  Here, we present 
interesting and instructive illustrations,  where we extract bound state eigenvalues from the known transmission coefficients 
$(T(E)=|t(E)|^2)$ for six  analytically solvable potential wells.
\end{abstract}
\maketitle
A one dimensional potential well that vanishes asymptotically can possess both bound  and scattering states. In  the former case, energy is negative and discrete:$E_n<0$, in the latter,  the energy is positive and continuous, where one can define a probability of  transmission which is called the transmission coefficient: $T(E)$. A square well and a Dirac delta well are simple examples [1-4]. Usually, textbooks discuss these two types of states separately.  Here we present the connection between negative energy bound states with the negative energy poles of $T(E)$.  This has been slightly touched upon  in the book by Merzbacher [1] but not without an error wherein the transmission coefficient  $T(E)$ at negative energy eigenvalues is mistakenly shown to become 1 in a Fig. 6.9, there. A potential barrier is a positive potential with a maximum that has only scattering states. If $V(x)=-V_0 f(x/a)$ represents a well of depth $V_0$, its inverted form $V(x)=U_0 f(x/a)$ gives a barrier of height $U_0$, here $V_0,U_0>0.$ In simple cases $f(\pm \infty)=0$, in more advanced cases $f(x/a)$ may diverge on one or both sides [2,3].

In textbooks, students are often asked to find the discrete set of energies at which  a square well or a barrier transmits fully [1,4]. This infinite set of  discrete energies also correspond to the bound state eigenvalues of an infinitely deep well but for an additive constant ($-V_0$). These discrete eigenvalues are different from the finite number of eigenvalues of a finite square well. 

For  simplicity, assume a single minimum potential well that vanishes asymptotically. For scattering of a particle with the energy $E>0$ from the left of  this potential, we require asymptotic form of the solution $\psi(x)$ of Schr{\"o}dinger equation
\begin{equation}
\frac{d^2\psi(x)}{dx^2}+\frac{2m}{\hbar^2}(E-V(x))\psi(x)=0.
\end{equation}
as 
\begin{equation}
\psi_<(x) \sim A(k) e^{ikx} +B(k) e^{-ikx}, \psi_>(x)\sim C(k) e^{ikx}.
\end{equation}
The reflection and transmission amplitudes are defined as 
\begin{equation}
t(k)=C(k)/A(k), \quad r(k)=B(k)/A(k).
\end{equation}
where the wave number $k=\sqrt{2mE}/\hbar$ and the energy $E>0$. The transmission and  reflection coefficients are defined as $T(E)=|t(k)|^2$ and $R(E)=|r(k)|^2$, respectively, such that $R(E)+T(E)=1$. For transition of $\psi(x)$ to  bound states at negative energies $(E<0)$, let $\kappa=\sqrt{-2m E}/\hbar$ such that $k=i\kappa$. We get
\begin{equation} 
\psi_<(x)\sim A(i\kappa) e^{-\kappa x} +B (i\kappa) e^{\kappa x},\psi_>(x)\sim C(i\kappa) e^{-\kappa x}.
\end{equation}
For bound states, to attain asymptotic convergence of $\psi(x)$ on both sides, we  set $A(\kappa)=0$ and this is how both $t(i\kappa)$ and $r(i \kappa)$ become infinity and we get discrete values of $\kappa (=\kappa_n)$.

Experts know that physical poles of $t(i\kappa)$ yield bound state eigenvalues. However, as mostly scattering coefficients and not the amplitudes are remembered, we propose  to extract bound states $E_n$ from  $T(E)$ or $R(E)$. Also if $\hat T(E)$ is known for a barrier of height $U_0>0$, we may change $U_0 \rightarrow -V_0$ to get $T(E)$ for the well. Even if the well diverges to $\pm \infty$,  the obtained $T(E)$ will suffice for obtaining the eigenvalues of bound states. One may see Illustrations 3 and 4.

It will be well to point out the  possible trivial/un-physical pole. In Eq. (4), $\kappa$ needs to be positive, otherwise $\psi(x)$ will diverge asymptotically. This is  despite the fact that $E=-\frac{\kappa^2\hbar^2}{2m}$ is insensitive to the sign of $\kappa$. Therefore, while extracting the poles, care must be taken to choose $\kappa_n>0$. Moreover, when the discrete spectrum is finite, this physical restriction defines the upper bound of the quantum number $n$. In this regard, see Illustrations 4 and 5.

One may get trivial/u-physical poles of complex amplitudes $t(k)$ or $r(k)$ having energy dependent uni-modular phase factor like $(1-i k)/(1+ik)$. Then on changing  $k \rightarrow i\kappa$, one will get a  pole $\kappa=1\implies E=-1$ (taking $\hbar^2=1=2m$), which being removable and independent of the potential parameters in $T(E)=|t|^2$ and $R(E)=|r|^2$ is inconsequential and hence un-physical. The reflection amplitude $r(k)$ of the exponential well $V(x)=-V_0 e^{-2|x|}$ [5] has  a factor of $\Gamma(ik)/\Gamma(-ik)$ whose modulus is 1, but by changing $k \rightarrow i\kappa$ and taking $\kappa=n$, we encounter an un-physical pole as $\Gamma(-n)=\infty, n\in I^+$ [6], This gives rise to  fake energy eigenvalues $E=-n^2$. Thus finding poles of $T(E)$ or $R(E)$ is more robust and meaningful. Also the complex transmission amplitude $t(k)$ may involve more advanced functions like gamma functions $\Gamma(z)$ of  complex variables but they get eliminated and give rise to simple form for $T(E)$, see  Illustration 6 below in this regard.

{\bf Illustration 1: Dirac Delta $V(x){=}-V_0 \delta(x), V_0{>}0$}

The transmission amplitude is given as   $t(k)=\hbar^2 k/(\hbar^2 k-im V_0)$ [4], whose pole is at $k=imV_0/\hbar^2$ giving the commonly known single bound state eigenvalue of this potential well at $E=k^2=-m V_0^2/(2\hbar^2)$. Similarly, $T(E)=2\hbar^2 E/(2\hbar^2E+m V_0^2)$, has a pole  at the same energy.

{\bf Illustration 2: Square well potential} 

For  this well $V(|x| \le a/2)=-V_0, V(|x|>a/2)=0$, whose depth is $V_0$ and width is $a$,
the commonly known eigenvalue equations for the even and odd parity states are given by
\begin{eqnarray} 
K a\tan(K a/2)=\kappa a ~~\text{and}~~  -K a\cot(K a/2) =\kappa a, \\ \nonumber K=\sqrt{2m(V_0+E)}/\hbar,~ \kappa=\sqrt{-2mE}/\hbar, E<0.
\end{eqnarray}
Let us convert (5) to $\sin$ and $\cos$ form respectively as
\begin{equation}
|\sin (K a/2)|=\frac{\kappa}{k_0}~ \text{and} ~ |\cos(Ka/2)|=\frac{\kappa}{k_0}  ~~, k_0=\frac{\sqrt{2mV_0}}{\hbar},
\end{equation}
$\tan(K a/2)$ should be positive and negative, respectively. This will be useful in sequel and such equations for eigenvalues of a square well  have been set up earlier for simplified graphical calculations in [7,8].
For a square well of depth $V_0$ and width $a$, the transmission amplitude is given by [1-4]
\begin{equation}
t(k)=\frac{e^{2ika}}{\cos (K a)-i\frac{k^2+K^2}{2k K} \sin (K a)}, E>0.
\end{equation}
For $E<0$, change $k \rightarrow i\kappa$, the poles of (7) are given by
\begin{equation}
\tan(K a)=\frac{2\kappa K}{K^2-\kappa^2}=\frac{2 \tan (K a/2)}{1-\tan^2 (K a/2)}.
\end{equation}
Solving it as an equation in $\tan(K a/2)$, we retrieve the energy eigenvalue equations (5). The transmission coefficient of a square barrier of width $a$ and height $U_0>0$ is commonly known [1-4] as 
\begin{eqnarray}
\hat T(E)=\frac{4\epsilon(1-\epsilon)}{4\epsilon(1-\epsilon)+\sinh^2(\alpha\sqrt{1-\epsilon})},\\ \nonumber \epsilon =\frac{E}{U_0}, \alpha=\frac{\sqrt{2mU_0}}{\hbar}a, E,V_0>0.
\end{eqnarray}
By changing $U_0 \to -V_0,~ \alpha \to ik_0 a, ~\epsilon =- \varepsilon$, we get $T(E)$ for the square well as
\begin{eqnarray}
T(E)=\frac{4\varepsilon(1+\varepsilon)}{4\epsilon (1+\varepsilon)+ \sin^2(k_0\sqrt{1+\varepsilon)}},\\ \nonumber \varepsilon =\frac{E}{V_0}, E, V_0 >0.
\end{eqnarray}
This also follows from (7) as $T(E)=|t(k)|^2$.
Now we look for the poles of $T(E)$ in (10), when $E<0$. The denominator gets factored in $\varepsilon $, the poles are given by
\begin{small}
\begin{eqnarray}
-\varepsilon{=}\frac{1}{2}(1\pm \cos(K a)){\implies} |\sin(K a/2)|=\sqrt{-\varepsilon}~\text{and}
\\ \nonumber ~~ |\cos(K a/2)|= \sqrt{-\varepsilon},
\end{eqnarray}
\end{small}
which is nothing but the energy eigenvalue equations given in (6) and equivalently in (5).
To re-emphasize, the poles given by two parts in (11) will give correct eigenvalues of the square well  only if their roots meet the criterion that $\tan(K a/2)$ is positive and negative, respectively. This is a slightly inconvenient aspect of square well for  extracting eigenvalues from $T(E)$ rather than $t(E)$ as done from Eq. (7). The following illustrations are free from this hassle.

A potential barrier that diverges to $-\infty$ on one side or both, upon inversion ($U_0\to -V_0$)  gives a well which diverges to $+\infty$ on respective sides, both $T$ and $R$ become un-physical yet they remain useful to extract discrete energies. The following two illustrations are for two such cases.

{\bf Illustration 3: Harmonic oscillator: $V(x)=\frac{1}{2}m \omega^2 x^2$:}

The transmission coefficient of the parabolic barrier $V(x)=V_0-\frac{1}{2}m \Omega^2 x^2$
is well known as Hill-Wheeler  formula [3,9]
\begin{equation}
\hat T(E)=\frac{1}{1+\exp\left(\frac{2\pi(V_0-E)}{\hbar \Omega}\right)}.
\end{equation}
If we put $\Omega =i\omega$ and $V_0=0$, the parabolic barrier becomes a well for which we have
\begin{equation}
T(E)=\frac{1}{1+\exp\left(\frac{2i\pi E}{\hbar \omega}\right)}.
\end{equation}
This being imaginary, it  is un-physical and  a  warning that a harmonic oscillator which is an inverted parabolic barrier does not admit scattering states, as it diverges to $+\infty$ on both sides of the well. However, (13) yields poles which are  the commonly known energy eigenvalues [1-4] of the harmonic oscillator.
\begin{eqnarray}
e^{2i\pi E/(\hbar \omega)}=-1=e^{(2n+1)i\pi},~ n=0,1,2,3,..\\ \nonumber \implies E_n=(n+1/2)\hbar \omega. 
\end{eqnarray}

{\bf Illustration 4: Morse oscillator, $V(x)=V_0[e^{2x/a}-2e^{x/a}]$}

This potential well [2,3] diverges to $+\infty$ on the right side, so it reflects a particle coming from left completely  without   transmission. But its corresponding barrier $V(x)=-U_0[e^{2x/a}-2e^{x/a}], U_0>0$ admits simple exact transmission coefficient [10] which is given for $E>0$ as
\begin{eqnarray}
\hat T(E)=\frac{1-e^{-4\pi \alpha}}{1+e^{2\pi(\beta-\alpha)}},~ \alpha=\sqrt{\frac{E}{\Delta}},~ \beta=\sqrt{\frac{U_0}{\Delta}},\\ \nonumber ~\Delta=\frac{\hbar^2}{2ma^2}.
\end{eqnarray}
By changing $U_0 \to -V_0 (\beta =i \eta (\eta=\sqrt{V_0/\Delta}))$,$ \alpha=i\gamma (\gamma=\sqrt{-E/\Delta}$), we get $T(k)$ which becomes non-real and un-physical,  nevertheless we get the poles as
\begin{small}
\begin{eqnarray}
e^{2i\pi(\eta-\gamma_n)}{=}{-}1{=}e^{\pm (2n+1)i\pi}\implies \gamma_n{=}[\eta-(n +1/2)]\\ \nonumber >0,\implies E_n=\gamma_n^2=-[\sqrt{V_0}-(n+1/2)\sqrt{\Delta}]^2,
\end{eqnarray}
\end{small}
which are the commonly known eigenvalues of the Morse potential. Since Morse potential well [2,3] admits only a finite number of eigenvalues , then $n=0,1,2,...<[\sqrt{V_0/\Delta}-1/2]$, here $[z]$ means the integer-part of $z$ .

{\bf Illustration  5: Soliton (Eckart) potential well}, $V(x)=-V_0 \text{sech}^2(x/a), V_0>0$ 

The transmission coefficient [2,3] for this barrier $V(x)=U_0 \text{sech}^2 (x/a)$ is given as
\begin{eqnarray}
T(E)=\frac{\cosh(2\pi \alpha)-1}{\cosh (2\pi \alpha)+\cos 2\pi\beta},~ \alpha=\sqrt{\frac{E}{\Delta}}, \\ \nonumber \beta =\sqrt{1/4-\frac{U_0}{\Delta}}.
\end{eqnarray}
Let us change $U_0 \to -V_0, ~ \alpha \to i \gamma (\gamma=\sqrt{-E/\Delta}),~ \beta= \eta= \sqrt{1/4+V_0/\Delta}$.We get
\begin{small}
\begin{eqnarray}
\cos(2\pi \gamma_n)=-\cos(2\pi \eta)=\cos(2\pi \eta-\pi) \implies 2\pi \gamma_n \\ \nonumber=\pm 2n \pi  \pm  (2\pi \eta -\pi)  \implies  \gamma_n=\eta-(n+1/2)]>0 \\ \nonumber \implies E_n=\gamma_n^2=-\Delta[\eta-(n+1/2)]^2.
\end{eqnarray}
\end{small}
In Eq. (16) and (18) the signs $\pm$ have to be chosen such that $\gamma_n>0$ $(\psi(x)\sim e^{-\gamma_n|x|})$ can decay when $x \to \infty$ and $n$ gets restricted  rightly  as these  potentials have a finite number of bound states. 

{\bf Illustration  6: Scarf II potential}

This exactly solvable potential is given by [11,12]
\begin{small}
\begin{equation}
V(x){=}\frac{\hbar^2}{2m}[(B^2{-}A^2{-}A)\text{sech}^2x{+}B(2A{+}1)\text{tanh}x{~}\text{sech}x].~(19)\nonumber
\end{equation}
\end{small}
Here $A$ and $B$ are essentially real and positive and $V(x)$ is Hermitian asymmetric potential well.
The very interesting transmission amplitude for (19) is given as [11]
\begin{small}
\begin{equation}
t(k){=}\frac{\Gamma({-}A{-}ik)\Gamma(1{+}A{}-ik)\Gamma(\frac{1}{2}{+}iB{-}ik)\Gamma(\frac{1}{2}{-}iB{-}ik)}{\Gamma(-ik)\Gamma(1+ik) \Gamma^2(1/2-ik)}~(20)\nonumber
\end{equation}
\end{small}
Noting that $\Gamma(-N)=\infty, N=0,1,2,3,...$, changing $k \rightarrow i\kappa$ the poles of $t(k)$ can be obtained as $-A-ik=-n \implies \kappa^{(1)}=(A-n)$.
The second $\Gamma(z)$, gives $\kappa^{(2)}=-n-A-1<0$, Third and fourth $\Gamma(z)$  gives $\kappa^{(3,4)}=(n+1/2)\pm iB$. Among these poles,  only $\kappa^{(1)}$ is physical. We get
\begin{equation}
\kappa^2_n=(A-n)^2{\implies}E_n=-\frac{\hbar^2}{2m}(A-n)^2.~~~~~~~~~~~~~~~~~~~(21)\nonumber
\end{equation}
These are exactly the eigenvalues obtained by the usual method of bound states [12],
Interestingly, we can find $T(E)=t(k)t^*(k)$ using (20) by writing $T(E)={\cal N}(E)/{\cal D}(E)$, where
\begin{small}
\begin{eqnarray}
{\cal N}(E){=}\Gamma(-z)\Gamma(-z^*) \Gamma(1{+}z)\Gamma(1{+}z^*)\Gamma(\frac{1}{2}{+}ix_1)\Gamma(\frac{1}{2}{-}ix_1)~(22) \nonumber \\  \Gamma(\frac{1}{2}{+}ix_2)\Gamma(\frac{1}{2}{-}ix_2), z=(A+ik), x_1=(B+k), x_2=(B-k)\nonumber
\end{eqnarray}
\end{small}
and 
\begin{small}
\begin{equation}
{\cal D}(E){=}\Gamma(-ik)\Gamma(ik)\Gamma(1{+}ik)\Gamma(1-ik)\Gamma^2(\frac{1}{2}-ik) \Gamma^{2}(\frac{1}{2}+ik)~(23) \nonumber
\end{equation}
\end{small}
Using the properties of $\Gamma$ functions [6]: $\Gamma(1+z)=z \Gamma(z), \Gamma(-z)\Gamma(z)=-\pi/z\sin (\pi z)$, $\Gamma(1/2+u)\Gamma(1/2-u)=\pi/\cos(\pi u)$. Further, after interesting manipulations with trigonometric and hyperbolic functions, we get
\begin{equation}
T(E)=\frac{\sinh^2 2\pi k}{(\cosh 2\pi k-\cos 2\pi A)(\cosh 2\pi k + \cosh2\pi B)}~~~(24) \nonumber
\end{equation}
One can easily see that when $k \rightarrow i\kappa$ in above, we get poles of $T(k)$ as $\cos 2 \pi \kappa=\cos 2\pi A \implies \kappa_n =(n \pm A)$
Choosing the $-$ sign and letting $0\le n< A$ ensure a finite number of bound state eigenvalues which are identically the same as in (21).

At research levels,  non-orthodox complex  (Parity-Time) potentials (invariant  under the joint action of $i\to -i, x \to -x$) [13]  and pseudo-Hermitian potentials [14]  are known to have real discrete spectrum. The extraction of eigenvalues from the transmission coefficient  has revealed a surprising and unnoticed presence of real discrete spectrum in a non-Hermitian potential, accidentally [15]. Interestingly, this model is neither PT-symmetric nor apparently pseudo-Hermitian. However,  the correctness of the claimed  spectrum has been tested by the other usual methods of finding real energy bound states. 

Lastly, we hope that  students will feel self-persuaded to extract
eigenvalues of bound states of more advanced and involved potential wells which have interesting analytic forms for transmission coefficients. These potentials are: Rosen-Morse [16], Ginocchio [17] and one more potential [18] which interpolates between Eckart and Morse potentials.  Even potential models which have $T(E)$ in terms of higher order functions (see [19]) are worth studying in this regard. The poles of reflection coefficient $R(E)$ can also be considered which are usually the same as that of $T(E)$. In more involved  cases, if they are not the same, then the common poles of the two will need to be considered.

\section*{\Large{Acknowledgement}}

We thank the physics trainees of 64th batch of Orientation Course of Engineering and Science (OCES) of the Human Resource Development Division (HRDD) of Bhabha Atomic Research Center for their keen interest in the course of Quantum Mechanics.

\end{document}